\documentstyle[12pt,a4,epsfig]{article}

\newcommand{\be}{\begin{equation}}
\newcommand{\ee}{\end{equation}}
\newcommand{\nn}{\nonumber}
\newcommand{\bea}{\begin{eqnarray}}
\newcommand{\eea}{\end{eqnarray}}
\newcommand{\eq}[1]{eq.~(\ref{#1})}

\newcommand{\gsim}{\ \rlap{\raise 2pt\hbox{$>$}}{\lower 2pt \hbox{$\sim$}}\ }
\newcommand{\lsim}{\ \rlap{\raise 2pt\hbox{$<$}}{\lower 2pt \hbox{$\sim$}}\ }

\newcommand{\matr}{\left( \begin{array}}
\newcommand{\ematr}{\end{array} \right)}

\def\beq{\begin{equation}}
\def\eeq{\end{equation}}
\def\bea{\begin{eqnarray}}
\def\eea{\end{eqnarray}}
\def\bq{\begin{quote}}
\def\eq{\end{quote}}
\def\ben{\begin{enumerate}}
\def\een{\end{enumerate}}
\def\nn{\nonumber}
\def\fr{\frac}

\def\ra{\rightarrow}

\def\dm{\Delta m}
\def\dtm{\Delta \tilde{m}}

\def\cp1sub{\setlength{\unitlength}{8pt}\begin{picture}(2,1)\mbox{\scriptsize CP} \end{picture}}                                 

\makeatletter
\if@twoside
\else \oddsidemargin 00pt \evensidemargin 00pt
\fi
\let\@eqnsel = \hfil

\expandafter\ifx\csname mathrm\endcsname\relax\def\mathrm#1{{\rm #1}}\fi
\@ifundefined{mathrm}{\def\mathrm#1{{\rm #1}}}{\relax}

\makeatother

\begin{document}

\begin{titlepage}
\begin{flushright}
CERN-TH/2001-032 \\
hep-ph/0102184
\end{flushright}

\vskip 1.5cm

\begin{center}
{\Large \bf      
MEDIUM EFFECTS FOR TERRESTRIAL AND
ATMOSPHERIC NEUTRINO OSCILLATIONS \par} \vskip 2.em
{\large         
{\sc M.C. Ba\~nuls$^1$, G. Barenboim$^{2}$ and J. Bernab\'eu$^{2,3}$
}  \\[1ex] 
{\it $^1$ IFIC, Centro Mixto Universitat de Val\`encia - CSIC} \\
{\it Edificio Investigaci\'on Paterna, E-46071 Val\`encia, Spain}\\
{\it $^2$ CERN - TH Division, \\
{CH-1211 Gen\`eve 23, 
Switzerland} \\
{\it $^3$ Departament de F\'\i sica Te\`orica, Universitat 
de Val\`encia}\\ 
{\it C/ del Dr. Moliner 50, 
E-46100 Burjassot (Val\`encia), Spain.} \\
\vskip 0.5em
\par} }

\vspace{2cm}

{\bf Abstract} \end{center}

Matter effects in neutrino propagation translate into effective parameters
for the oscillation and fake CP- and CPT-odd quantities, even
in a scenario, such as $\Delta_{12}=0$, where no genuine CP  
violation is present.
This fact seems to impose severe restrictions on the determination 
of intrinsic parameters of the system from long-baseline experiments.
We show, however, that the resonance in the effective mixing
$\tilde{\theta}_{13}$ can be observed for a certain range of baselines.
This provides a way to measure the vacuum mixing angle 
 $\theta_{13}$ and the sign of $\Delta m_{23}^2$ from atmospheric
neutrinos, using a detector with energy resolution and charge
discrimination.

\par
\vfill
\noindent
February 2001 
\end{titlepage}

\section{Introduction} 
Present evidence of neutrino oscillations in atmospheric neutrinos
provides the range of values for the oscillation phase governed
by $\Delta m_{23}^2 $ to be \cite{uno}
\be
1.5 \;10^{-3} {\mbox{eV}}^2 \leq
\; \mid\Delta m_{23}^2\mid \;\leq 5 \; 10^{-3} {\mbox{eV}}^2 ,
\ee
and the neutrino mixing in the
corresponding sector, i.e. $\theta_{23}$, to be  near maximal.
On the contrary, the solution to the solar neutrino behaviour still 
presents several alternatives, the most favoured one 
\cite{dos} being the LAMSW solution
for $\theta_{12}$ and $\Delta m_{12}^2 $, that would allow,
if connected by a non-vanishing $\theta_{13}$ to the other sector,
the possibility of CP violation in leptonic physics. 
Up to now, $\theta_{13}$  has been bounded by CHOOZ~\cite{tres}
with a value
\bea
\sin^2\theta_{13} \leq 0.05.
\eea

The studies around terrestrial long-baseline experiments and neutrino
factories have precisely this objective in mind, the exploration
of CP violation in the lepton case
\cite{cuatro, cinco, cincob}, a possibility open only 
for non-degenerate neutrino masses and
non-vanishing mixings on the neutrino
propagation. With this aim, the automatic
self-inclusion of a new actor, the matter effect, 
is somehow seen as undesirable, as a background 
which could avoid the 
determination of the intrinsic properties of the neutrinos.

In this letter we will take a different attitude and show that, 
under appropriate conditions, matter effects bring the connecting 
mixing angle $\theta_{13}$ into the game, even when the sector $(1,2)$
is irrelevant. In fact, we will consider terrestrial and
atmospheric neutrino oscillations, with baselines such that, in 
a good approximation, the oscillating phase $\Delta_{12}$, defined
as  $\Delta_{12} =  \Delta m_{12}^2 L /(4E)$, can be neglected.

For  $\Delta_{12}=0$, the mixing in the sector $(1,2)$ is inoperative
and there is no room for {\bf{genuine}} CP violation. However, the
neutrino interaction with an asymmetric medium leads to (fake)
CP-odd and CPT-odd non-vanishing quantities, which are 
by themselves a clean indicator of the effects to be searched for. 
These quantities, if
non-vanishing, 
distinguish neutrinos 
from antineutrinos and their sign automatically indicates
that of $\Delta m_{23}^2$.

General arguments \cite{cinco}
teach us that the  difference in the
survival probabilities for neutrinos and antineutrinos needs a
CPT-odd origin, so this observable, particularly for $\nu_\mu$ versus
$\bar{\nu}_\mu$ where there are good prospects to distinguish the
charge, will be of interest. There is a drawback, however, for this
proposal, if the medium effect acts as a perturbative modification. 
The dominant (``allowed'') vacuum oscillations would take place in
the $(2,3)$ sector and there would be no possibility
for the small $\theta_{13}$ to show up.
In this perturbative scenario, 
the alternative proposal has been
to emphasize the ``forbidden'' appearance channel $\nu_e 
\rightarrow \nu_\mu $ \cite{cincob, seis, siete}
appropriate for neutrino factories and candidate to
the search for genuine CP violation.

Another path can be to study whether the interaction
of neutrinos with matter can generate an
observable resonant situation in
the effective mixing $\tilde{\theta}_{13}$, which naturally 
incorporates information on $\theta_{13}$. 
The question then arises, of
which the conditions of observability of such a resonance are,
given the strict bounds on $\theta_{13}$.

There seems to be  a fatalistic result, 
a sort of no-go theorem, stating that  the maximum
in the effective mixing is compensated by a minimum in the
oscillation 
and only the product is measured. 
This cancelation seems to be also operative for the
T-violating probabilities \cite{sieteb}.
Although it is true
that this compensation operates in many physical situations of
practical interest, and we will confirm this, it cannot always be
true, because the mixing is independent of the baseline whereas
the oscillating phase depends linearly on it. We will study 
the conditions under which the resonant $\tilde{\theta}_{13}$
transition can be made visible in the disappearance channel,
opening the door to an appreciable effect in the CPT-odd 
$\nu_\mu$ survival probability and accessible in atmospheric
neutrino experiments.

The plan of this paper is as follows. In Section 2 we develop the relevant
amplitudes for 3-family neutrino oscillations in matter of constant
density, in the approximation that $\Delta_{12} =0 $. A resonant behaviour
appears for the effective mixing  $\tilde{\theta}_{13}$ in matter with
a width proportional to $\theta_{13}$.
Section 3 gives the behaviour with energies outside the resonance region.
The interference pattern, 
for small $ \theta_{13}$, is shifted from the
vacuum $L/E$ behaviour to an additional term proportional to $L$ 
and independent of $E$. The prospect is to determine 
$ \theta_{13}$ from the forbidden channel probabilities 
$\nu_e \longrightarrow \nu_\mu$ and 
$\bar{\nu}_e \longrightarrow \bar{\nu}_\mu$. 
In Section 4 we study the conditions on $L$ under which the 
resonance can be made visible: the
disappearance channel $\nu_\mu \longrightarrow \nu_\mu$ then becomes the
simplest and most spectacular one. 
The CPT-odd asymmetry is 
able not only to show the resonance effects but also
to give information on both the magnitude of $ \theta_{13}$
and the sign of $\Delta m_{23}^2$. From the resonance effect in both
$\nu_\mu \longrightarrow \nu_\mu$ and
$\nu_e \longrightarrow \nu_\mu$ channels, we estimate the charge asymmetry
in atmospheric neutrinos.
In Section 5 we summarize
our conclusions.

\section{Basics of three-neutrino oscillations in matter}

The effective hamiltonian that describes the 
time evolution of neutrinos 
in matter can be written in the flavour basis as \cite{ocho}
\beq
H=\fr{1}{2 E}\left\{U
\left(\begin{array}{c c c} 0&&\\&\dm_{12}^2&\\&&\dm_{13}^2\end{array}\right)
U^+  +
\left(\begin{array}{c c c} a&&\\&0&\\&&0\end{array}\right) \right\},
\eeq
where $a=G\sqrt{2}N_e 2E$ represents matter effects from the 
effective potential of electron-neutrinos with electrons, and 
$U$ is the flavour mixing matrix in vacuum (PDG representation 
\cite{nueve}),
\beq
U=\left (
\begin{array}{c c c}
c_{12} c_{13} & s_{12} c_{13} & s_{13} e^{-i\delta}\\
-s_{12}c_{23}-c_{12}s_{23}s_{13}e^{i\delta} & 
c_{12}c_{23}-s_{12}s_{23}s_{13}e^{i\delta} & s_{23}c_{13}\\
s_{12}s_{23}-c_{12}c_{23}s_{13}e^{i\delta} &
-c_{12}s_{23}-s_{12}c_{23}s_{13}e^{i\delta} & c_{23}c_{13}
\end{array}
\right ).
\eeq

For a baseline $L$, the evolution of neutrino states is given by
\beq
\nu(L)=S(L) \nu(0),
\eeq
with
\beq
S(L)=e^{-i H L}
\eeq
for constant matter density. The corresponding effective hamiltonian
for antineutrinos is obtained by $U \rightarrow U^*$ and
$a \rightarrow -a$.

If the hamiltonian can be separated into two pieces as
$H=H_0+H_1$,
where $H_0$ can be exactly solved and $H_1$ can be treated in perturbation 
theory, then $S(L)=S_0(L)+S_1(L)$ and
$S_0(L)=e^{-i H_0 L}$ gives the lowest order transition amplitudes.

For $\mid\dm_{12}^2\mid \;
\ll \; \mid\dm_{13}^2\mid$, and $a$ in a region of energies
such that it is comparable to $ \mid \dm_{13}^2 \mid$, 
we choose to solve exactly
\bea
H_0&=&\fr{1}{2 E}\left\{U
\left(\begin{array}{c c c} 0&&\\&0&\\&&\dm_{13}^2\end{array}\right)
U^{+}+
\left(\begin{array}{c c c} a&&\\&0&\\&&0\end{array}\right) \right\} \nn\\
&=&\fr{1}{2E}\tilde{U}
\left(\begin{array}{c c c} 0&&\\&\dtm_2^2&\\&&\dtm_3^2\end{array}\right)
\tilde{U}^{+},
\eea
where $\dtm^2$ describes the energy-level spacings in matter and
$\tilde{U}$ the effective mixings. One realizes that the matter effect
breaks the degeneracy.
Therefore,
\beq
S_0(L)=\tilde{U}
\left(\begin{array}{c c c} 0&&\\&e^{-i\fr{\dtm_2^2}{2 E}L}&\\
&&e^{-i\fr{\dtm_3^2}{2 E}L} \end{array}\right)
\tilde{U}^{+}.
\eeq

The exact diagonalization of $H_0$ yields effective mass differences 
and mixings in matter, in the limit $\dm_{12}\ra0$, 
but with no approximation taken on $a$.
We get the following result
\beq
\tilde{U}=\left(\begin{array}{ccc}
0&\fr{e^{-i\delta}}{n_2}(l_2-c_{13}^2)&\fr{e^{-i\delta}}{n_3}(l_3-c_{13}^2)\\
-c_{23}&\fr{1}{n_2}s_{23}s_{13}c_{13}&\fr{1}{n_3}s_{23}s_{13}c_{13}\\
s_{23}&\fr{1}{n_2}c_{23}s_{13}c_{13}&\fr{1}{n_3}c_{23}s_{13}c_{13}
\end{array}\right),
\eeq
for the mixings in terms of the vacuum parameters, where 
we have defined
\beq
n_2=\sqrt{l_2^2-2l_2c_{13}^2+c_{13}^2}; \quad
n_3=\sqrt{l_3^2-2l_3c_{13}^2+c_{13}^2}, 
\eeq
and 
\bea
l_2\equiv\fr{\dtm_2^2}{\dm_{13}^2}&=&
\fr{1}{2}\left[1+\alpha-
\sqrt{1+\alpha^2-
2\alpha\cos(2 \theta_{13}) }\right];\nn\\
l_3\equiv\fr{\dtm_3^2}{\dm_{13}^2}&=&
\fr{1}{2}\left[1+\alpha+
\sqrt{1+\alpha^2-
2\alpha\cos(2 \theta_{13}) }\right],
\eea
which give the effective mass differences.
We have introduced the dimensionless
parameter $\alpha\equiv\fr{a}{\dm_{13}^2}$
for the sake of simplicity. The ordering of levels in matter is
(1,2,3) for the hierarchical case of $\Delta m_{13}^2 > 0 $ and
$\alpha < 1$.

The effective mixing matrix $\tilde{U}$ 
is independent of $\theta_{12}$ and $\delta$, if 
the latter is appropriately rotated away.
The matrix $\tilde{U}$  can be expressed also 
in the PDG form.
To do so, we change
\beq
\tilde{U} \ra \tilde{U}\left( \begin{array}{ccc}
1&&\\&e^{i \delta}&\\&&1
\end{array}
\right ).
\eeq
By comparing with previous expressions, we may establish the
correspondence between the effective mixing angles, $\tilde{\theta}_{ij}$,
and the vacuum parameters, getting
\bea
\tilde{c}_{12}=0, \hspace{3cm} |\tilde{s}_{12}|=1, \nn \\
\tilde{c}_{23}=c_{23}, \hspace{3cm} \tilde{s}_{23}=s_{23}, \nn \\
\tilde{c}_{13}=\fr{l_2-c_{13}^2}{n_2}, \hspace{2cm}
\tilde{s}_{13}=\fr{l_3-c_{13}^2}{n_3},
\eea
up to signs. The vanishing mixing in matter $\tilde{c}_{12}=0$
is a consequence of the degeneracy $\Delta_{12}=0$ in vacuum and
says that 
the lowest mass eigenstate in matter contains no 
electron-neutrino flavour component.
This result is the ingredient that avoids genuine CP
violation in matter, even if one has three non-degenerate effective
masses.

Transition amplitudes for $\nu_\alpha \ra \nu_\beta$
in the $\dm_{12}=0$ case are given 
by $S_0$ matrix elements, which can be written
\bea
A(\alpha\ra\beta;L)&=&S_0(L)_{\beta \alpha} \\
&=&
\delta_{\beta \alpha}+\tilde{U}_{\beta 2}\tilde{U}_{2\alpha}^+
\left(e^{-i\fr{\dtm_2^2}{2 E}L}-1 \right)
+\tilde{U}_{\beta 3}\tilde{U}_{3\alpha}^+
\left(e^{-i\fr{\dtm_3^2}{2 E}L}-1 \right). \nn
\eea
From this expression we may calculate all probabilities,
\bea
P\left(\nu_e \rightarrow \nu_e \right) &=& 1 - \sin^2(2\tilde{\theta}_{13})
 \sin^2\left[ \tilde{\Delta}_{13} \right] \\
P\left(\nu_\mu \rightarrow \nu_e \right) &=& s_{23}^2 
\sin^2(2\tilde{\theta}_{13})
 \sin^2\left[\tilde{\Delta}_{13} \right] \\
P\left(\nu_\tau \rightarrow \nu_e \right) &=& c_{23}^2 
\sin^2(2\tilde{\theta}_{13})
 \sin^2\left[\tilde{\Delta}_{13}\right] \\
P\left(\nu_\mu \rightarrow \nu_\mu \right)&=& 1 - s_{23}^4
\sin^2(2\tilde{\theta}_{13})\sin^2\left[\tilde{\Delta}_{13}\right] -
2 s_{23}^2 c_{23}^2 \left\{ 1 - \cos\left[\Delta_{13} (1 + \alpha) \right]
\cos\left[\tilde{\Delta}_{13}\right] 
\right. \nonumber \\
&+& \left. \cos(2\tilde{\theta}_{13})\sin\left[\Delta_{13} (1 + \alpha) \right]
\sin \left[\tilde{\Delta}_{13} \right] \right\} 
 \\
P\left(\nu_\tau \rightarrow \nu_\mu \right)&=&
s_{23}^2 c_{23}^2 \left\{ 2 - 2\cos\left[\Delta_{13} (1 + \alpha) \right]
\cos \left[\tilde{\Delta}_{13} \right] -
\right. 
 \sin^2(2\tilde{\theta}_{13})
 \sin^2\left[ \tilde{\Delta}_{13}  \right]  \nonumber \\
&+&\left. 2 \cos(2\tilde{\theta}_{13}) \sin\left[\Delta_{13} (1 + \alpha) \right]
\sin \left[ \tilde{\Delta}_{13} \right] \right\}
 \\
P\left(\nu_\tau \rightarrow \nu_\tau \right)&=& 1 - c_{23}^4
 \sin^2(2\tilde{\theta}_{13})
 \sin^2\left[ \tilde{\Delta}_{13} \right]  - 
2 s_{23}^2 c_{23}^2 \left\{ 1 - \cos\left[\Delta_{13} (1 + \alpha) \right]
\cos \left[\tilde{\Delta}_{13}\right] 
\right.  \nonumber \\
&+& \left. \cos(2\tilde{\theta}_{13})\sin\left[\Delta_{13} (1 + \alpha) \right]
\sin \left[\tilde{\Delta}_{13}\right] \right\} 
\eea
where 
\bea
\tilde{\Delta}_{13}\equiv  \Delta_{13}
\sqrt{1+\alpha^2-
2\alpha\cos(2 \theta_{13}) },
\; \; \; {\mbox{with}}\; \; \;
 \Delta_{13} \equiv \frac{\Delta m_{13}^2 L}{4 E} .
\eea 
and
\beq
\sin^2(2\tilde{\theta}_{13})=4\fr{s_{13}^2c_{13}^2}
{1+\alpha^2-
2\alpha\cos(2 \theta_{13})}.
\label{sin2t}
\eeq
The probabilities for time-reversal-conjugated transitions satisfy
$(\alpha \neq \beta)$
\bea
P\left(\nu_\beta \rightarrow \nu_\alpha \right)=
P\left(\nu_\alpha \rightarrow \nu_\beta \right).
\eea
as they are even functions of the baseline $L$ \cite{cinco}.

In order to get the corresponding expressions for
antineutrinos, we must change $a \ra -a $, i.e. $\alpha \ra -\alpha $. 
The effect of such a change in the probabilities comes from the
different relative sign between mass and matter terms in $H_0$.
The same effect can be achieved by changing not the sign of $a$
but that of $\Delta m_{13}^2$, i.e. by considering a different 
hierarchy for mass eigenstates in vacuum.

Now we analyse the resonance. Without taking any limit on $s_{13}$, Eq.
(\ref{sin2t}) can be written as 
\bea
\sin^2(2\tilde{\theta}_{13}) &=& \frac{ 4 s_{13}^2 c_{13}^2}
{(\alpha - \cos 2\theta_{13})^2 + 4 s_{13}^2 c_{13}^2 } \nonumber \\
&&\\
&=& \frac {4 s_{13}^2 c_{13}^2 \left( \frac{ \Delta m_{13}^2}
{\tilde{a}} \right)^2 }{ 
\left(  E - \cos 2\theta_{13} \frac{\Delta m_{13}^2}
{\tilde{a}} \right)^2 + 4 s_{13}^2 c_{13}^2 
\left( \frac{ \Delta m_{13}^2}
{\tilde{a}} \right)^2 }, \nonumber 
\eea
where
\bea
\tilde{a}=2 \sqrt{2} G_F N_e.
\eea
From here we obtain the resonant energy, given by
\bea
E_R = \cos (2\theta_{13}) \frac{\Delta m_{13}^2}
{\tilde{a}},
\eea
and the width
\bea
\Gamma = 2 \sin(2\theta_{13}) \frac{\Delta m_{13}^2}
{\tilde{a}}.
\eea
If interpreted in terms of the variable $\alpha$, the resonant
parameters are given by
\bea
\alpha_R&=& \cos(2 \theta_{13});
\label{res}
\nonumber \\
\Gamma_{\alpha}&=&4 s_{13} c_{13}.
\eea

It is important to stress that the resonant energy is not sensitive
to $s_{13}$, for small $\theta_{13}$,
 as it varies like the cosine, but is a measure
of $\Delta m_{13}^2$. On the other hand, the resonance width
depends linearly 
on $\theta_{13}$ and can be a useful tool to measure it.

An inspection of Eqs.(15)--(20) for the probabilities in different
channels points out 
that there are contributions from both 
the imaginary part squared of
 the resonant amplitude, $\sin^2(2\tilde{\theta}_{13})$, and the 
interference with the real part, $\cos(2\tilde{\theta}_{13})$. 
On top of the resonance, $\sin^2(2\tilde{\theta}_{13})=1$,
independent of $s_{13}$, and $\cos(2\tilde{\theta}_{13})=0$.

For $\Delta m^2_{13} > 0$, the resonance appears only for neutrinos,
whereas for $\Delta m^2_{13} < 0$ it would show up
only for antineutrinos.

\section{Outside the resonance}
When we are far from the resonance ($\alpha \simeq 1$), 
so that $s_{13}$ is small with respect to $\alpha -1$, 
up to quartic terms in $s_{13}$ 
for the probabilities we can neglect the effects of the width,
\bea
\sqrt{1+\alpha^2-
2\alpha\cos(2 \theta_{13})}  \simeq \mid \alpha - 1 \mid
\eea
and therefore the transition probabilities can be written as
\bea
P\left(\nu_e \rightarrow \nu_e \right) &=& 1 - \frac{4 s_{13}^2}
{(1 - \alpha)^2}
 \sin^2\left[ \Delta_{13} (1 - \alpha )\right] \\
P\left(\nu_\mu \rightarrow \nu_e \right) &=& s_{23}^2 \frac{4 s_{13}^2}
{(1 - \alpha)^2}
 \sin^2\left[ \Delta_{13} (1 - \alpha )\right] \\
P\left(\nu_\tau \rightarrow \nu_e \right) &=& c_{23}^2 \frac{4 s_{13}^2}
{(1 - \alpha)^2}
 \sin^2\left[ \Delta_{13} (1 - \alpha )\right] \\
P\left(\nu_\mu \rightarrow \nu_\mu \right) &=& 1- s_{23}^2 \frac{4 s_{13}^2}
{(1 - \alpha)^2}
 \sin^2\left[ \Delta_{13} (1 - \alpha )\right] - 4 s_{23}^2 c_{23}^2
\sin^2 \left[ \Delta_{13} \right]
\label{mm}\\
P\left(\nu_\tau \rightarrow \nu_\mu \right) &=& 4 s_{23}^2 c_{23}^2 
\sin^2 \left[ \Delta_{13} \right] - s_{23}^2 c_{23}^2 \frac{4 s_{13}^2}
{(1 - \alpha)^2}
 \sin^2\left[ \Delta_{13} (1 - \alpha )\right]  \\
P\left(\nu_\tau \rightarrow \nu_\tau \right) &=& 1 - c_{23}^4 
\frac{4 s_{13}^2}
{(1 - \alpha)^2}
 \sin^2\left[ \Delta_{13} (1 - \alpha )\right] - 4 s_{23}^2 c_{23}^2 
\sin^2 \left[ \Delta_{13} \right].
\eea
As was pointed out already in \cite{siete}, the forbidden transition
$P\left(\nu_\mu \rightarrow \nu_e \right)$ 
presents an interference pattern which,
 in addition to the vacuum
$\Delta_{13} \sim L/E$ dependence of the oscillation phase, has 
an energy independent phase
shift induced by matter
as, $ \Delta_{13}\alpha  \sim   L$, 
thus providing  a possibility of quantifying matter effects. This 
phase shift is due to a purely quantum-mechanical effect with potentials 
and corresponds to an analogous to the Minkowski-rotated form of the 
Aharonov--Bohm experiment \cite{diez}. Instead of space interference, 
one has here flavour interference; the interferometer becomes the 
mixing matrix, the optical path difference the value of $\Delta_{13}$,
  and the energy-independent effective potential $a/(2E)$.

As widely recognized in the literature, the 
$\nu_e \rightarrow \nu_\mu$ transition probability 
provides   a very good measurement of $s_{13}$. 
A very high flux  is needed
to be sensitive to its forbidden character.
In the limit $\Delta_{13} (1-\alpha) \ll 1$, 
the $\ \sin^2\left[ \Delta_{13} (1 - \alpha )\right] $ factor is compensated
by the $1/(1-\alpha)^2$ enhancement. Thus, matter effects would 
remain small for either small $L/E$, or $\alpha$ 
far away from the resonance, or both.
Under these conditions, the (fake) CP-odd probability would be suppressed.
On the contrary, once $L/E$ is increased, the appearance
neutrino probability is
dramatically increased (for $\Delta m_{13}^2 >0$) respect to
that for propagation in vacuum, whereas the antineutrino probability 
becomes much smaller. To illustrate this feature quantitatively, we give 
in Figs. 1 and 2 the $\nu_e \rightarrow \nu_\mu$
probabilities for $L$ = 3000 and 7000 km, respectively.
(For $L$ = 700 km the effects are not appreciable.)
The three curves, dotted, solid and dashed, correspond
in each figure to neutrino,
vacuum and antineutrino probabilities.
Notice that, besides the change in magnitude, there is a shift in the 
oscillation pattern, as imposed by an attractive or repulsive potential.
In all these cases, we have taken $\Delta m_{23}^2 = 3 \; 10^{-3}$ eV$^2$,
$\tilde{a}= 2.8 \; 10^{-13} $eV and $\theta_{13}=0.23$.
Under these conditions the mixing resonance has parameters 
$E_R =9.6$ GeV with half-width $\Gamma /2 = 4.7$ GeV. An inspection
of these results shows no special role of the resonance,
except for $L=7000$ km (see next section). In fact, important matter
effects appear at $L=3000$ km for both neutrinos and antineutrinos,
indicating that they are non-resonant.

Contrary to the forbidden channel $\nu_e \rightarrow \nu_\mu$
(or $\nu_e \rightarrow \nu_\tau$ or the dissappearance
$\nu_e \rightarrow $\hspace{-.5cm} / $\; \nu_e$), the allowed channel
$\nu_\mu \rightarrow \nu_\tau$ or the survival probability
$\nu_\mu \rightarrow \nu_\mu$ have contributions coming from the
interference with the real part $\cos(2\tilde{\theta}_{13})$
of the amplitude, besides the imaginary part squared,
$\sin^2(2\tilde{\theta}_{13})$. Outside the resonance, the first 
contribution dominates and all the physics is controlled by the (2,3)
sector, without any room for appreciable matter effects. Needless
to say, $s_{13}$ plays no role in these cases. It remains to be seen
whether there are new features around the resonance,
which could allow $s_{13}$ to show up.

\section{On the observability of the resonance}

The main difficulty for the observation of resonant effects in the
``connecting'' mixing $\tilde{\theta}_{13}$ comes from the product
compensation, in measurable quantities, of the effective mixing
resonance and the oscillation factor.
Therefore, to induce appreciable effects of the resonance we have 
to impose an overlap of the mixing peak with a maximum of the oscillation
factor.
If this is possible, is a way of escaping the washing out
effect that was thought to be unavoidable.

The maximal mixing, $\sin^2(2\tilde{\theta}_{13})=1$, is reached on
top of the resonance, corresponding to a value of $\alpha $ equal to
$\alpha_R $, Eq.(\ref{res}).
On the other hand, the first maximum of the oscillation term
is reached when the oscillation phase,
$ \Delta_{13} \sqrt{1+\alpha^2- 2\alpha\cos(2 \theta_{13}) }$,
takes the value $\pi /2$, 
that corresponds to a value $\alpha_{\mbox{\tiny {max}}}$.
By imposing the condition that both maxima coincide, i.e.
that $\alpha_R =\alpha_{\mbox{\tiny {max}}}$,
we obtain the baseline that maximizes the observability
of the resonant effect,
\bea
L_{\mbox{\tiny {max}}}= \frac{ 2 \pi}{\tilde{a} \tan(2 \theta_{13})}.
\label{max}
\eea 
Notice that $L_{\mbox{\tiny {max}}}$ is independent of $\Delta m_{13}^2$,
which determines the resonant energy, and it is inverse to $\theta_{13}$.
The condition to avoid the cancellation of the resonant effect
by the oscillation does not need to be so restrictive. 
One could allow a separation between the two
maxima of one resonance half-width,
given in terms of $\alpha$ by $\Gamma_\alpha /2$, Eq.(\ref{res}),
 and still expect to observe the overlap.
The corresponding baseline for this less restrictive condition
would be
\bea
L_{\mbox{\tiny {min}}} \simeq \frac{L_{\mbox{\tiny {max}}}}
{\sqrt{2}}.
\eea
For $\tilde{a}$ and $\theta_{13}$ as taken in the previous section,
$L=7000$ km approximately satisfies this condition.

In Fig. 3  we give the survival probabilities
$\nu_\mu \rightarrow \nu_\mu$ for 
$L$ = 7000 km.
Again the three curves in each figure correspond to neutrino, vacuum
and antineutrino probabilities. For the short baselines, the
physics is dominated by the (2,3) sector without any appreciable
matter effect even at $L$ = 3000 km, but 
there appears a spectacular change 
of regime for $L=$ 7000 km, in which the resonance becomes
apparent. 
In this way, matter effects (through the resonance) are only
important in one channel: neutrinos (antineutrinos) for
$\Delta m_{23}^2 > 0\; (<0)$.

An impressive plateau around the resonance is
the signal expected for the CPT-odd asymmetry, with
a sign opposite to that of $\Delta m_{23}^2 $.
We give in Fig. 4 the (fake) CPT-odd 
asymmetry for the discussed case of $\theta_{13} = 0.23$ and for
$\theta_{13} = 0.15$. We conclude that, for appropriate $L$, the
muon-neutrino 
survival probability is sensitive to the resonance effect in matter
and its CPT-odd asymmetry provides a measure of the 
connecting mixing $\theta_{13}$ in vacuum.

To quantify the implication of such an indicator for the 
zenith angle effect in atmospheric neutrino oscillations, we 
proceed in the following way. We convolute both, the $\nu_\mu$
survival probability (Fig. 3) and the $\nu_\mu$ appearance probability
from $\nu_e$ (Fig. 2) with the corresponding $\nu_\mu$ and
$\nu_e$ atmospheric fluxes \cite{atm} and with the $\nu_\mu$
cross section in matter \cite{nueve}. Similarly for antineutrinos.
Notice that the difference of behaviour for $\nu_\mu$ and
$\bar{\nu}_\mu$ comes now from both, the matter effect and the convolution.
The observed muon charge asymmetry is plotted in Fig. 5 as a function of
energy in the interesting resonance region. Our results are 
obtained    for
$L \; \sim$ 7000 km, within a zenith angle resolution of 5$^o$. We
conclude that, even for this minimum value of $L$ in the sense
of Eq.(37), the four values of the set ($\theta_{13}, \; \pm\mid
\Delta m_{23}^2\mid$) can be distinguished. Needless to say,  one
can amplify the effect going to values of $L$ closer to
$L_{\mbox{\tiny {max}}}$ (Eq.(36)), leading to a better
sensitivity for smaller values of $\theta_{13}$.

\section{Conclusions} 
\label{conclusions}
In this paper we have explored the medium effects in neutrino 
oscillations for baselines appropriate to terrestrial or
atmospheric neutrinos. The analysis has been made in the approximation
$\Delta_{12} =0$ and it has been mainly applied to both the 
forbidden appearance channel $\nu_e \ra \nu_\mu$ and the
survival probability  for $\nu_\mu \ra \nu_\mu$. The manifestation
of the matter effects has been presented in terms of the
fake CP-odd and CPT-odd asymmetries, respectively. These observables
are sensitive to the connecting mixing angle $\theta_{13}$ in 
magnitude and to the sign of $\Delta m_{23}^2$.

We have analysed the change of regime in going from a short baseline
of 700 km to a long baseline of 7000 km. For the latter, we are entering 
into a manifestation of the resonance present in the 
effective mixing $\sin^2(2\tilde{\theta}_{13})$.

The forbidden appearance $\nu_e \ra \nu_\mu$ probabilities, which are very
sensitive to $s_{13}$, show already at  $L=$ 3000 km very important matter
effects, 
which are non-resonant, with phase shifts of opposite sign for neutrinos
and antineutrinos. The CP-even probability, relevant for detectors without
charge discrimination, still sees appreciable matter effects, more 
apparent in oscillation phase shifts than in magnitude. Of course,
the sensitivity to $s_{13}^2$ in magnitude is there, and a variation 
in $s_{13}$ does not affect the oscillation pattern. At $L=$ 7000 km, matter 
effects become resonance-dominated and affect neutrinos
both in magnitude and phase (for $\Delta m^2_{23} >0$).

Contrary to the transition in regime discussed for $\nu_e \ra \nu_\mu$,
the disappearance channel $\nu_\mu \ra \nu_\mu$ only sees matter effects 
from a baseline above $L \sim $ 7000 km, i.e. when the resonance shows up. 
Even at $L=$ 3000 km, one cannot induce appreciable  medium effects.
This is understood: outside the resonance, the physics is here 
dominated by the ``allowed'' sector (2,3), which is not sensitive to
interactions with matter. Once the resonance in 
 $\sin^2(2\tilde{\theta}_{13})$ operates, medium effects 
appreciably modify the magnitude for the neutrino channel but not for
antineutrinos, where no resonance appears for $\Delta m^2_{23} >0$. 
The corresponding CPT-odd asymmetry, shown in Fig. 4,
is sensitive to the connecting mixing $s_{13}$ in its magnitude and
its sign distinguishes the sign of $\Delta m^2_{23}$.

The calculated muon charge asymmetry originated from both
 $\nu_\mu \ra \nu_\mu$  and 
 $\nu_e \ra \nu_\mu$  shows important effects in the resonance
region. Fig. 5 shows that we are able to distinguish the values
of $\theta_{13}$, and the sign of $\Delta m^2_{13}$. Notice that this sign
is not automatically translated into the sign of the asymmetry, 
as it was in Fig. 4. We have estimated that a 10 kT detector with 
energy resolution and charge discrimination
can reach a few percent accuracy in the measured 
asymmetry in one year of data taking.

\subsection*{Acknowledgements}
We are grateful to Pilar Hern\'andez for interesting discussions
and to Solveig Skadhauge for computing help. This work has been 
supported by CICYT, Spain, under Grant AEN99-0692.

\newpage

\begin{figure}[!ht]
  \begin{center}
  \epsfig{file=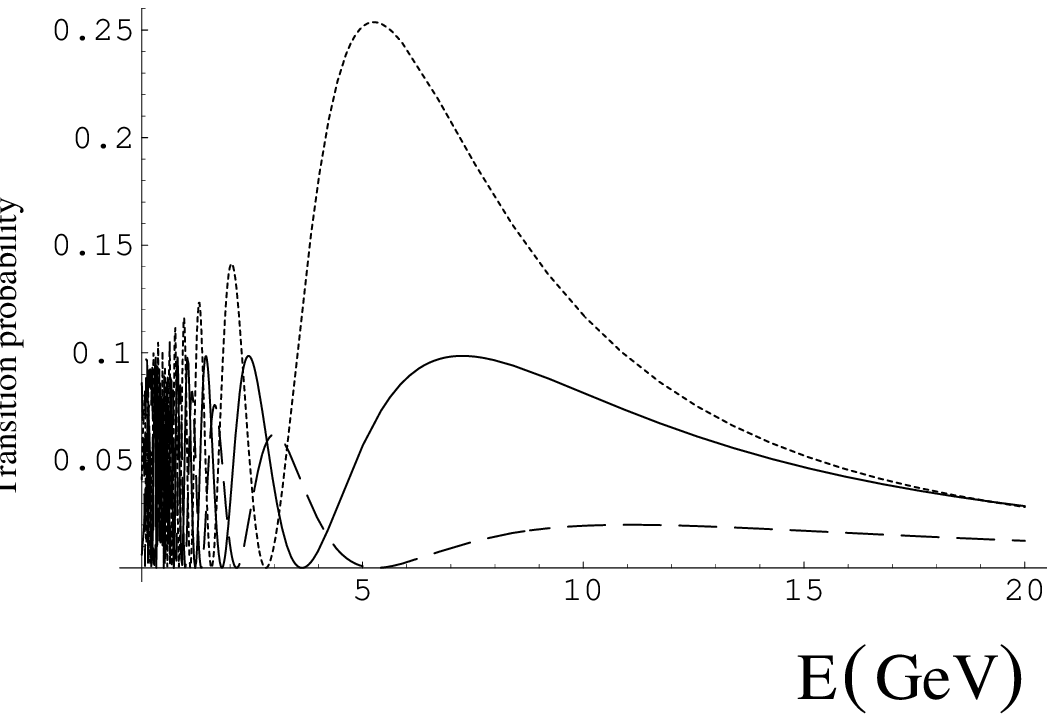,width=8cm}
\caption{ $\nu_e \ra \nu_\mu$ transition probability
for neutrinos (dotted), vacuum (solid) and antineutrinos (dashed)
for a baseline $L=$ 3000 km. } 
  \end{center}
\end{figure}

\begin{figure}[!ht]
  \begin{center}
  \epsfig{file=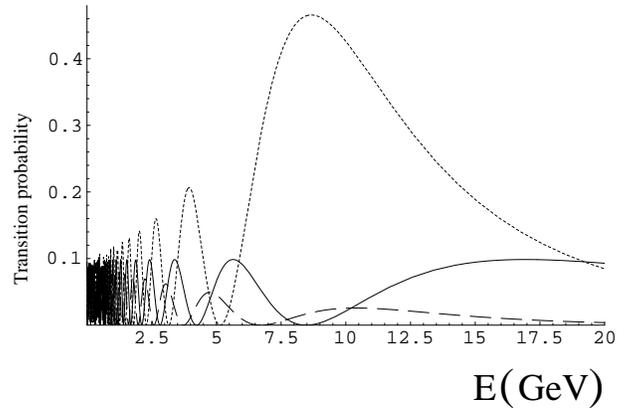,width=8cm}
\caption{Same as Fig. 1 for $L=$ 7000 km.} 
  \end{center}
\end{figure}

\begin{figure}[!ht]
  \begin{center}
  \epsfig{file=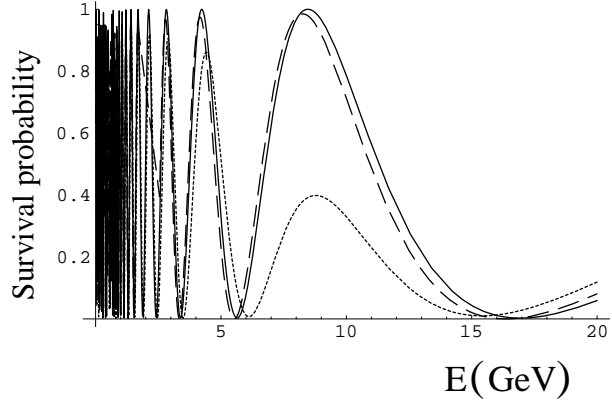,width=8cm}
\caption{Muon neutrino survival probability 
for neutrinos (dotted), vacuum (solid) and antineutrinos (dashed)
for a baseline $L=$ 7000 km.} 
  \end{center}
\end{figure}

\begin{figure}[!ht]
  \begin{center}
  \epsfig{file=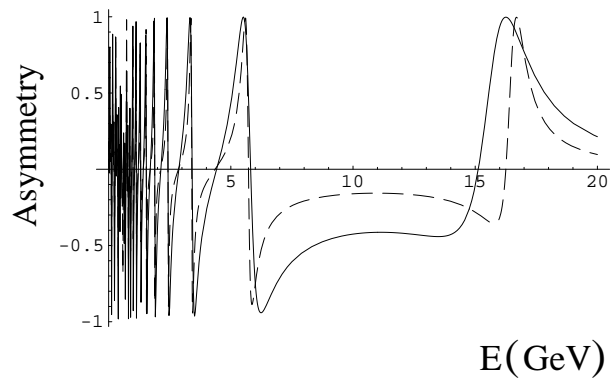,width=8cm}
\caption{(Fake) CPT-odd  asymmetry 
for muon neutrinos and
$L=$ 7000 km. Solid line for $\theta_{13} = 0.23$,
dashed line for $\theta_{13} = 0.15$.} 
  \end{center}
\end{figure}

\newpage

\begin{figure}[!ht]
  \begin{center}
  \epsfig{file=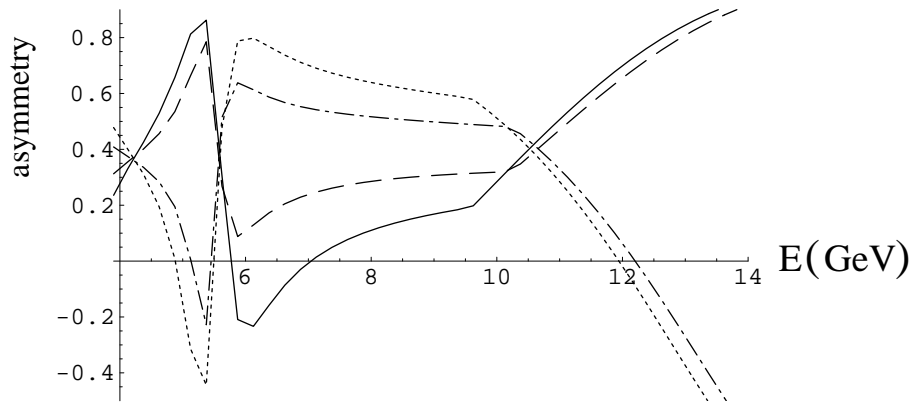,width=12cm}
\caption{Muon charge asymmetry for $\theta_{13} = 0.23$ and
sign($\Delta m_{23}^2$)= + (solid), $\theta_{13} = 0.15$ and
sign($\Delta m_{23}^2$)= + (dashed), $\theta_{13} = 0.23$ and
sign($\Delta m_{23}^2$)= - (dotted) and $\theta_{13} = 0.15$ and
sign($\Delta m_{23}^2$)= - (dash-dotted)
} 
  \end{center}
\end{figure}

\end{document}